\documentclass[usenatbib,usegraphicx,onecolumn]{mn2e}

\usepackage{epsfig}
\usepackage[usenames]{color}

\begin{document}

\title[Transit Surveys in {star clusters}] {On the
  potential of transit surveys in {star clusters}:
  Impact of correlated noise and radial velocity follow-up}

\author[S.\ Aigrain et al.]{Suzanne~Aigrain$^{1,2}$ and 
  Fr{\' e}d{\' e}ric~Pont$^{3}$ \\
  $^{1}$Institute of Astronomy, University of Cambridge, Madingley
  Road,
  Cambridge, CB3 0HA, United Kingdom \\
  $^{2}$School of Physics, University of Exeter, Stocker Road, Exeter, EX4 4QL, United Kingdom \\
  $^{2}$Observatoire Astronomique de l'Universit{\' e} de Gen{\` e}ve,
  51, chemin des Maillettes, CH-1290 Sauverny, Switzerland}
\date{Accepted \ldots Received \ldots; in original form\ldots}

\maketitle

\begin{abstract}
  We present an extension of the formalism recently proposed by Pepper
  \& Gaudi to evaluate the yield of transit surveys in
  {homogeneous stellar systems}, incorporating the
  impact of correlated noise on transit time-scales on the
  detectability of transits, and simultaneously incorporating the
  magnitude limits imposed by the need for radial velocity follow-up
  of transit candidates. New expressions are derived for the different
  contributions to the noise budget on transit time-scales and the
  least-squares detection statistic for box-shaped transits, and their
  behaviour as a function of stellar mass is re-examined. Correlated
  noise that is constant with apparent stellar magnitude implies a
  steep decrease in detection probability at the \emph{high} mass end
  which, when considered jointly with the radial velocity
  requirements, can severely limit the potential of otherwise
  promising surveys in {star clusters. However, we
  find that small-aperture, wide field surveys may detect hot Neptunes
  whose radial velocity signal can be measured with present-day
  instrumentation in very nearby ($<100$\,pc) clusters.}
\end{abstract}

\section{Introduction}

Open Clusters have long been used as laboratories to test our
understanding of star formation and stellar evolution, as each
contains a (sometimes large) sample of stars with relatively
well-known and common properties (age, composition, environment) but
spanning a wide range of masses. For the same reasons, since the
discovery of the first extra-solar planet around a Sun-like star in
the field just over a decade ago \citep{mq95}, the possibility of
discovery extra-solar planets in open clusters has been a tantalising
goal.

Open Clusters tend to be relatively distant, and their members
relatively faint compared to the field stars usually targeted by
radial velocity surveys. Most of the projects searching for
extra-solar planets in Open Clusters therefore employ the transit
technique, which is particularly well suited to dense stellar
environments and has the additional advantage of providing a direct
measurement of the planet to star radius ratio. Recent or ongoing
transit surveys in Open Clusters include the UStAPS (the University of
St Andrews Planet Search, \citealt{shl+03,bhm+05,hck+05}), EXPLORE-OC
\citep{vls+05}, PISCES (Planets in Stellar Clusters Extensive Search,
\citealt{mss+05,mss+06}), STEPSS (Survey for Transiting Extra-solar
Planets in Stellar Systems, \citealt{bgd+06}) and the \emph{Monitor}
project \citep{ahi+07}.

Understanding the factors that affect the yield of such a survey is
vital not only to maximise its detection rate, but also to enable the
interpretation of the results of the survey, including in the case of
non-detections, in terms of constraints on the incidence and parameter
distributions of planetary companions. Recently, \citet{pg05a}
(hereafter PG05a) introduced an analytical formalism to estimate the
rate of detection of exo-planets via the transit method in stellar
systems. One particularly interesting result is the fact that the
probability that transits of a given system in a given cluster are
detectable, if they occur, is a very slowly varying function of
stellar mass in the regime where the photometric performance is
dominated by the source photon noise, but drops sharply with stellar
mass in the background-dominated regime. This implies that the number
of detections to be expected from a given survey is roughly
proportional to the number of stars with source photon counts above
the sky background level in that survey.

In a follow-up paper, \citet{pg05b} (hereafter PG05b) applied the
aforementioned formalism to young open clusters, showing that transit
surveys focusing on these systems have the potential to detect
transiting Neptune and even Earth-sized planets, by making use of the
fact that low-mass stars are relatively bright at early ages, and that
their smaller radius gives rise to deeper transits for a given planet
radius. This opens up the tantalising possibility of detecting
transits of terrestrial planets from the ground, and what is more of
doing so in young systems, where one might obtain particularly
interesting constraints on the formation and evolution of extra-solar
planets \citep{ahi+07}.

The formalism of PG05a assumes that the photometric errors on each
star and in each observation are independent of each other (i.e.\ that
the noise is white). However, an \emph{a posteriori} analysis of the
detection threshold of the OGLE transit survey in the light of their
RV follow-up observations of OGLE candidate transits \citep{pbm+05}
demonstrated that the effective detection threshold is significantly
higher than that expected for white noise only, suggesting that
correlated noise on transit time-scales might be present in OGLE light
curves. \citet*{pzq06} (hereafter PZQ06) since developed a set of
methods for evaluating the amount of correlated noise on transit
time-scales in the light curves of transit surveys, and applied them
to the OGLE light curves to show the latter do indeed contain
correlated noise at the level of a few mmag. Similar analysis of light
curves from other transit surveys (see e.g.\ \citealt{scc+06} and
\citealt{iia+07}, and \citealt{pon07} for an overview) has since shown
that {they} are also affected by correlated noise at a
similar level. Because correlated noise does not average out as more
observations of a given transit event are obtained, as white noise
does, it is generally correlated noise which dominates over white
noise in determining the detectability of transits
{around all but} the faintest stars in a given field
survey. A number of effects can give rise to correlated noise,
including seeing-dependent contamination of the flux measured for a
given star by flux from neighbouring stars, pointing drifts combined
with flat-fielding errors, and imperfect sky subtraction -- some or
all of which may be important in a given survey depending on the
telescope/instrument combination used and the observing strategy.

Photometry alone does not allow the mass of the companion causing the
transits to be ascertained, and radial velocity measurements are thus
generally needed to confirm the planetary nature of a transit
event. As pointed out by PG05a, this effectively imposes an apparent
magnitude limit for transit detections to be confirmable, as accurate
radial velocity measurements of faint stars are extremely expensive in
terms of large telescope time. 

As {noise that is correlated on transit timescales} reduces the detectability of transits around the
brightest stars in a given survey, but the need to perform radial
velocity follow-up implies that only around the brightest stars can
transits be confirmed, both effects must be incorporated in
the scaling laws used to estimate the number of detections expected
from a given survey. The present paper attempts to do this by
extending the formalism of PG05a to include correlated noise and by
translating the magnitude limit imposed by radial velocity follow-up
into a cluster-specific mass limit.

Section~\ref{rednoise} briefly sketches out the basics of the
formalism of PG05a and describes how one or more additional noise terms
representing correlated noise terms can be incorporated in it. The
impact of these modifications on the noise budget on transit
time-scales and on the transit detection probability as a function of
mass are investigated in Section~\ref{impact}. Considerations external
to the transit search itself, including radial velocity follow-up, are
introduced in Section~\ref{addcons}. Finally, the practical
implications of the resulting formalism for Open Cluster transit
searches are briefly explored in Section~\ref{examples}.

\section{Introducing red noise terms}
\label{rednoise}

\subsection{Overall formalism}

This formalism is described in detail in PG05a, and only its major
characteristics are sketched out here, so as to allow the
modifications implied by the presence of correlated noise to be made
clear.

PG05a compute the number of transiting planets with periods between $P$
and $P+dP$ and radii between $r$ and $r+dr$ that can be detected
around stars with masses between $M$ and $M+dM$ in a given stellar
system as
\begin{equation}
\frac{d^3N_{\rm det}}{dM~dr~dP} =  N_{\star} f_p \frac{d^2p}{dr~dP} 
\mathcal{P}_{\rm tot} (M,~P,~r) \frac{dn}{dM}.
\end{equation}
where $N_{\rm det}$ is the number of detected transiting planets,
$N_{\star}$ is the total number of stars in the system, $d^2p/dr~dP$
is the probability that a planet around a star in the system has a
period between $P$ and $P+dP$ and a radius between $r$ and $r+dr$,
$f_p$ is the fraction of stars in the system with planets,
$\mathcal{P}_{\rm tot}(M,~P,~r)$ is the probability that a planet of
radius $r$ and orbital period $P$ will be detected around a star of
mass $M$, and $dn/dM$ is the mass function of the stars in the system,
normalised over the mass range corresponding to $N_{\star}$.

Following \citet{gau00}, PG05a separate $\mathcal{P}_{\rm tot}
(M,~P,~r)$ into three factors:
\begin{equation}
\mathcal{P}_{\rm tot} (M,~P,~r) = \mathcal{P}_{\rm tr} (M,~P) 
\mathcal{P}_{\rm S/N} (M,~P,~r) \mathcal{P}_{\rm W} (P)
\end{equation}
$\mathcal{P}_{\rm tr}$ is the probability that a planet transits its
parent star, $\mathcal{P}_{\rm S/N}$ is the probability that, should a
transit occur during a night of observing, it will yield a
signal-to-noise ratio (${\rm S/N}$) that is higher than some threshold
value, and $\mathcal{P}_{\rm W}$ is the window function that describes
the probability that more than one transit will occur during the
observations.

PG05a's expression for the transit probability is used without
modification
\begin{equation}
P_{\rm tr} = \frac{R}{a} = 
\left( \frac{4\pi^2}{G} \right)^{1/3} M^{-1/3} RP^{-2/3}
\end{equation}
where $R$ is the star radius and $a$ the orbital distance.

The ${\rm S/N}$ of a set of transits is ${\rm S/N} =
\left(\Delta\chi^2_{\rm tr}\right)^{1/2}$, where $\Delta\chi^2_{\rm
  tr}$ is the difference in $\chi^2$ between a constant flux and a
boxcar transit fit to the data. {PZQ06 give the general
  expression:
  \begin{equation}
    \label{eq:dc2s_gen}
    \Delta \chi^2_{\rm tr} = \frac{d^2}{\sigma^2_d} = \frac{d^2\,n^2}{\sum C_{ij}}
  \end{equation}
  where $d$ is the transit depth, $\sigma_d$ is the uncertainty on the
  transit depth, $n$ is the number of in-transit data points and $C$
  is the covariance matrix the in-transit flux
  measurements\footnote{{One can show that the
      estimate of $d$ which minimises the $\chi^2$ of the fit is the
      inverse-variance weighted average of the in-transit
      flux-measurements, and $\sigma_d$ is thus the uncertainty on
      this average.}}. If the noise is uncorrelated, the non-diagonal
  elements of $C$ are zero, and $\sum C_{ij}=\sum_i\sigma^2_i$ where
  $\sigma_i$ is the uncertainty on the $i^{\rm th}$ flux
  measurement. Additionally, if this uncertainty is constant, i.e.\
  $\sigma_i=\sigma_{\rm w}$, Equation~(\ref {eq:dc2s_gen}) further reduces
  to:
  \begin{equation}
    \label{eq:dc2s_w}
    \Delta \chi^2_{\rm tr} = n \left( \frac{\delta}{\sigma_{\rm w}} \right)^2,
  \end{equation}
  which, for single transits, is equivalent to Equation~(4) in PG05a.
  Note that the notation adopted here matches that of PZQ06, and thus
  differs that of PG05a,b. In particular, the symbols $N_{\rm tr}$ and
  $n$, used in PG05a,b to represent the number of in-transit points
  and the number of transits respectivtely, are inverted here. We also
  use $\sigma_0$ where PG05a,b used $\sigma$, and $d$ where they used
  $\delta$.}
  
\subsection{{Modifying the detection statistic to
    account for red noise}}

In an attempt to account for the saturation of the rms.\ noise level
that is seen at the bright end of all transit surveys, PG05a introduced
in their Section~4.3 thqe concept of a \emph{minimum observational
  error} $\sigma_{\rm sys}$, which is added in quadrature to the error
contribution $\sigma_{\rm phot}$ from the sky and source photon noise
to give the error estimate $\sigma_{\rm ind}$ for each data point:
\begin{equation}
\sigma_{\rm w} = ( \sigma_{\rm phot}^2 + \sigma_{\rm sys}^2 )^{\frac{1}{2}}.
\end{equation}
This expression for $\sigma_{\rm w}$ is then simply inserted into
Equation~\ref{eq:dc2s_w}).

However, detailed analysis of the light curves of various transit
surveys (PZQ06, \citealt{pon07}) shows that they systematically
contain noise that is correlated on transit timescales (2--3\,h for a
Hot Jupiter transit), i.e.\ the non-diagonal elements of the
covariance matrix are non-zero. {As a result,
  $\sigma_d$ no longer decays as $n^{-1/2}$ as expected for
  uncorrelated (white) noise. PZQ06 propose a single parameter
  description of the covariance, assuming the noise can be separated
  into purely uncorrelated (white) and purely correlated (red)
  components, the former decaying as $n^{-1/2}$ but the latter
  independent of $n$:
  \begin{equation}
    \label{eq:sigtr}
    \sigma_d^2 = \frac{\sigma^2_{\rm w}}{n} + \sigma^2_{\rm r}
  \end{equation}
  where $\sigma_{\rm w}$ and $\sigma_{\rm r}$ reresent the white and
  red noise components respectively. This single parameter description
  of the correlated noise assumes that the degree of correlation
  remains unchanged on all timescales up to the maximum transit
  duration. It is equivalent to approximating the covariance matrix
  with $C_{ii}=\sigma_0^2 \equiv \sigma_w^2+\sigma_r^2$ in the
  diagonal, $C_{ij}=\sigma^2_r$ for two data points in the same
  transit, and $C_{ij}=0$ otherwise (see Section~\label{sec:mult} for
  the treatment of multiple transits).}

{There is evidence that the correlation timescale in
  transit survey light curves is finite \citep{gds+06}. If this
  correlation timescale is shorter than the maxium transit duration,
  the above expression would underestimate the significance of
  long-duration transit events. This does not appear to be the case
  for light curves analysed by PZQ06 and \citet{pon07}, where the
  noise remains correlated up to 3\,h timescales. Nevertheless, it is
  interesting to investigate the impact of finite correlation
  timescales through a simple example. We consider a transit with a
  depth of 1\% lasting 2\,h and observed with 15\,min time sampling,
  i.e. $n=8$. In the white noise only case, if $\sigma_{\rm
    w}=2$\,mmag, $\sigma_d=0.71$\,mmag and $\Delta\chi^2_{\rm
    tr}=200$. If correlated noise is present, with $\sigma_{\rm
    r}=1$\,mmag, the single parameter approximation gives
  $\sigma_d=1.22$\,mmag and $\Delta\chi^2_{\rm tr}=67$. If on the
  other hand the noise is correlated only over timescales up to 1\,h
  or 4 data points, i.e.\ $C_{ij}=0$ for $|i-j| \geq 4$,
  $\sigma_d=1.10$\,mmag and $\Delta\chi^2_{\rm tr}=82$. In general,
  even if the characteristic correlation timescale of the noise is
  shorter than a transit duration, we expect the single parameter
  correlated noise approximation adopted here to give an estimate of
  the transit significance that is much nearer to the true value than
  that obtained with the white noise approximation.}

{We therefore adopt Equation~(\ref{eq:sigtr}) for what
  follows.} The white noise is assumed to be equal to the photon noise
and modelled as of contributions from the source and the sky
background:
\begin{equation}
  \label{eq:sigw}
  \sigma^2_{\rm w} = 
  \sigma_{\rm source}^2 + \sigma_{\rm back}^2 = 
  \frac{N_{\rm s} + N_{\rm b}}{N_{\rm s}^2},
\end{equation}
where $N_{\rm s}$ and $N_{\rm b}$ are the number of photons from the
source and the sky detected in the photometric aperture. 

PZQ06 found that the distribution of the rms.\ of OGLE light curves
over a typical transit time-scale of 2.5\,h is consistent with a
constant red noise level of $\sigma_{\rm sys}\sim 3$\,mmag,
independent of apparent magnitude. Processing the light curves with a
systematics removal algorithm such as Sys-Rem reduces $\sigma_{\rm
  sys}$ to $\sim 1.5$\,mmag for the best objects. The work of the
International Space Science Institute (ISSI) working group on
transiting planets \citep{pon07} has shown that similar values are
also typical of other surveys, with a correlated noise value of $\sim
1.5$\,mmag for the best objects. We therefore adopt $\sigma_{\rm sys}
\sim 1.5$\,mmag throughout the following calculations, which would
correspond to very good ground-based
photometry. {While it is theoretically possible to
  reduce the level of correlated systematics further, this value is
  used because it is considered representative of the leading surveys
  currently in operation.}

In addition to this systematics term, a red noise component
proportional to the white noise level (as a function of magnitude) is
present in some surveys. This dominates over the systematics term in
the domain where background photon noise dominates the white noise and
is thus likely to be somehow associated with background
subtraction. We therefore label it $\sigma_{\rm sub}$. For the
purposes of the present calculations, it is modelled as a term
proportional to the background noise:
\begin{equation}
\label{eq:sigsub}
\sigma_{\rm sub} = k \sigma_{\rm b} 
\end{equation}
For the purposes of the present work we assume $k=0.2$. This is the
kind of values the ISSI team found for the correlated noise in the HAT
and SuperWASP surveys. Most cluster surveys, such as the University of
St Andrews Planet Search \citep{shl+03,bhm+05,hck+05}, EXPLORE--OC
\citep{vls+05}, STEPSS \citep{bgd+06}, PISCES
\citep{mss+05,hsg+05,mss+06} or Monitor \citep{ahi+07,iia+07}, have
`better spatial sampling, and lower values of $k$ might therefore be
expected to apply, though preliminary analysis of test light curves
indicates that $k\sim0.2$ is also appropriate, if not an
underestimate, for at least some of these surveys. In any case, this
value is used here to illustrate the effects of noise of this type
when it dominates the overall noise budget.

The overall red noise budget is thus
\begin{equation}
\label{eq:sigr}
\sigma^2_{\rm r} = 
\sigma_{\rm sys}^2 + \sigma_{\rm sub}^2 = \sigma^2_{\rm sys} + 
\frac{k^2 N_{\rm b}}{N_{\rm s}^2},
\end{equation}

\subsection{{Multiple transits}}

As correlated noise does not average out over transit time-scales, but
does average out over repeated transit events, it is particularly
important to consider the repeatability of transits in the detection
process when one suspects correlated noise might dominate. In an
appendix, PG05a derived an expression for $\mathcal{P}_{\rm S/N}$ for
multiple transits{, which is based on the equation
\begin{equation}
\label{eq:dc2n}
\Delta_{\rm tr}^2 ({\rm multiple~transits}) = 
N_{\rm tr} ~ \Delta_{\rm tr}^2 ({\rm single~transits}) = 
N_{\rm tr} \frac{d^2}{\sigma^2_d}
\end{equation}
where $N_{\rm tr}$ is the number of observed transits. This equation
remains valid in the presence of correlated noise provided there is no
correlation over long timescales (similar to the planet's orbital
period). In PG05a, $P_{\rm W}$ has to be calculated separately for
each value of the number of transits. This assumes that the number of
data points in each observed transit is the same, and in practice one
must therefore choose a minimum value for the number of data points in
a partially observed transit above which that transit contributes to
$\mathcal{P}_{\rm W}$, and below which it does not.}

{PZQ06 provide a general formula which accounts for
  the number of data points in each observed transit:
  \begin{equation}
    \Delta_{\rm tr}^2 ({\rm multiple~transits}) = 
    n_{\rm tot}^2 \frac{d^2}{\sum_{k=1}^{N_{\rm tr}}n^2_k \mathcal{V}(n_k)}
  \end{equation}
  where $n_{\rm tot}=\sum_k=1^{N_{\rm tr}}$ is the total number of
    in-transit points and
  \begin{equation}
    \mathcal{V}(n_k) \equiv 
    \frac{1}{n_k^2}\sum^{n_k \times n_k~{\rm block}} C_{ij}
  \end{equation}
  is the noise integrated over the $k^{\rm th}$ observed transit.
  $\mathcal{P}_{\rm W}$ then becomes a multi-dimensional quantity
  dependent on not only $N_{\rm tr}$ but for each $N_{\rm tr}$, on the
  set of $n_k$. As with the PG05a formalism, it must be evaluated
  numerically.}

{In the present work, we make the assumption of
  homogeneous phase coverage, which allows us to ignore differences
  between $n_k$ for the different transits, and enables us to
  (roughly) estimate the number of transits observed as a function of
  period given the time sampling and survey duration. This can then be
  incorporated} into Equation~(\ref{eq:dc2n}) directly, therefore
  alleviating the need to compute $P_{\rm W}$ separately. One can
  approximate $N_{\rm tr}$ as
\begin{equation}
\label{eq:n}
N_{\rm tr} = \frac{t_{\rm tot}}{P} = \frac{N_{\rm n} \, t_{\rm night}}{P}
\end{equation}
where $t_{\rm tot}$ is the total time spent on target, which is the
product of the number of nights $N_{\rm n}$ and the average duration
of a night $t_{\rm night}$, and $P$ is orbital
period. {Reality diverges strongly from the
  homoegeneous phase coverage assumption close to harmonics of the
  daily interruptions in the observations, but it follows the global
  $1/P$ trend (see Fig.~1 of PG05a).}

\section{Impact on the noise budget and detection statistic}
\label{impact}

\subsection{Noise budget on transit time-scales}
\label{sec:sigmatr}

Useful insights regarding the dominant noise sources, and how to
mitigate those that have the largest impact on the transit detection
performance, can be gained by exploring the dependency of the the
various noise components on the stellar parameters. We start from the
following expressions, given by PG05a, for $N_{\rm s}$, $N_{\rm b}$
and {$n$} (recalling that $n$ is called $N_{\rm tr}$ in PG05a):
\begin{equation}
\label{eq:ns}
N_{\rm s} = \frac{L_{X,\odot} 10^{-0.4A_X}}{4 \pi d^2} t_{\rm exp} 
\pi \left(\frac{D}{2}\right)^2 \left( \frac{M}{M_\odot}\right)^{\beta_X}
= N_{{\rm s},\odot} \left( \frac{M}{M_\odot}\right)^{\beta_X},
\end{equation}
\begin{equation}
\label{eq:nb}
N_{\rm b} = S_{{\rm sky},X}\Omega t_{\rm exp} \pi \left(\frac{D}{2}\right)^2,
\end{equation}
and
{
\begin{equation}
\label{eq:ntr}
n = \sqrt{1-b^2} \frac{R_\odot}{\delta t} 
\left(\frac{4 P}{\pi G M_\odot}\right)^{1/3} 
\left( \frac{M}{M_\odot}\right)^{\alpha-\frac{1}{3}} =
 \sqrt{1-b^2}~n_{{\rm eq},\odot} \left( \frac{M}{M_\odot}\right)^{\alpha-\frac{1}{3}} 
\end{equation}}
where $M$ is the stellar mass; $\alpha$ is the index of the (power-law)
mass-radius relation; $\beta_X$ is the index of the (power-law)
mass-luminosity relation in the filter $X$ under consideration; $D$ is
the telescope aperture; $t_{\rm exp}$ is the exposure time; $d$ is the
distance to the cluster; $A_X$ is the extinction to the cluster. PG05a
adopt the distance-dependent extinction law {$A_I=0.5 (d/{\rm
    kpc})$}; $L_{X,\odot}$ is the Sun's photon luminosity in the filter
of interest, which we compute, following PG05a, as
\begin{equation}
\label{eq:lsun}
L_{X,\odot} = \frac{8 \pi^2 c R^2_{\odot} \lambda^{-4}_{X,c} \Delta\lambda_X}
                       {\exp \left( hc / \lambda_{X,c} k T_\odot \right) -1},
\end{equation}
where $\lambda_{X,c}$ and $\Delta\lambda_X$ are the filter central
wavelength and FWHM respectively;
$S_{{\rm sky},X}$ is the sky photon flux per unit solid
  angle;
$\Omega$ is the effective area of the seeing disk, which we
  compute, following PG05a, as
\begin{equation}
\label{eq:omega}
\Omega = \frac{\pi}{\ln 4} \theta^2_{\rm see}
\end{equation}
where $\theta_{\rm see}$ is the FWHM of the PSF;
$b$ is the impact parameter of the transit;
$\delta t$ is the interval between consecutive measurements. In
  PG05a,
\begin{equation}
\delta t = t_{\rm exp} + t_{\rm read}
\end{equation}
where $t_{\rm read}$ is the readout time, which can be generalised to
include any time spent off-target; $N_{{\rm s},\odot}$ is the number
of source photons in the aperture for a solar-mass star;
{$n_{{\rm eq},\odot}$} is the number of points in an
  equatorial transit for a solar-mass star.

Substituting for $N_{\rm s}$ and $N_{\rm b}$ from
Equations~(\ref{eq:ns}) and (\ref{eq:nb}) into
Equation~(\ref{eq:sigw}) gives
{
\begin{equation}
\label{eq:sigw2}
\sigma_{\rm w}^2 = \frac{1}{N_{{\rm s},\odot}} 
\left( \frac{M}{M_\odot}\right)^{-\beta_X}
\left[1 + C_2 \left( \frac{M}{M_\odot}\right)^{-\beta_X} \right].
\end{equation}}
where {we} have introduced, following PG05a,
\begin{equation}
\label{eq:c2}
C_2 = \frac{4 \pi d^2 S_{{\rm sky},X}\Omega}
               {L_{X,\odot} 10^{-0.4A_X}} 
\end{equation}
which is the ratio of sky to source flux in the aperture for a solar
mass star. Taking $\sigma_{\rm r}$ from Equation~(\ref{eq:sigr}),
$\sigma_{\rm w}$ from Equation~(\ref{eq:sigw2}) and {$n$} from
Equation~(\ref{eq:ntr}) and substituting into
Equation~(\ref{eq:sigtr}),
\begin{equation}
\label{eq:sigtr2}
\sigma_d^2 = \frac{\left( 1 - b^2 \right)^{-1/2}}
{N_{{\rm s},\odot}n_{{\rm eq},\odot}}
\left( \frac{M}{M_\odot}\right)^{\frac{1}{3}-\alpha-\beta_X}
\left[1 + C_2 \left( \frac{M}{M_\odot}\right)^{-\beta_X} \right]
+ \sigma_{\rm sys}^2 + 
\frac{k^2 C_2}{N_{{\rm s},\odot}} \left( \frac{M}{M_\odot} \right)^{-2\beta_X}.
\end{equation}
{To simplify this expression we introduce two new constants
\begin{equation}
\label{eq:c4}
C_4 = N_{{\rm s},\odot}~n_{{\rm eq},\odot}
\end{equation}
which is the square of the background subtraction component for a Sun-like star, and 
\begin{equation}
\label{eq:c5}
C_5 = \frac{k^2 C_2}{N_{{\rm s},\odot}}
\end{equation}
which is the total number of source photons collected during an
equatorial transit for a Sun-like star. (Note that $C_3$ is defined in
PG05a but not used here.) Equation~(\ref{eq:sigtr2}) then becomes
\begin{equation}
\label{eq:sigtr3}
\sigma_d^2 = \frac{\left( 1 - b^2 \right)^{-1/2}}{C_4}
\left( \frac{M}{M_\odot}\right)^{\frac{1}{3}-\alpha-\beta_X}
\left[1 + C_2 \left( \frac{M}{M_\odot}\right)^{-\beta_X} \right]
+ \sigma_{\rm sys}^2 + C_5 \left( \frac{M}{M_\odot} \right)^{-2\beta_X}.
\end{equation}}

\begin{figure*}
\centering\epsfig{file=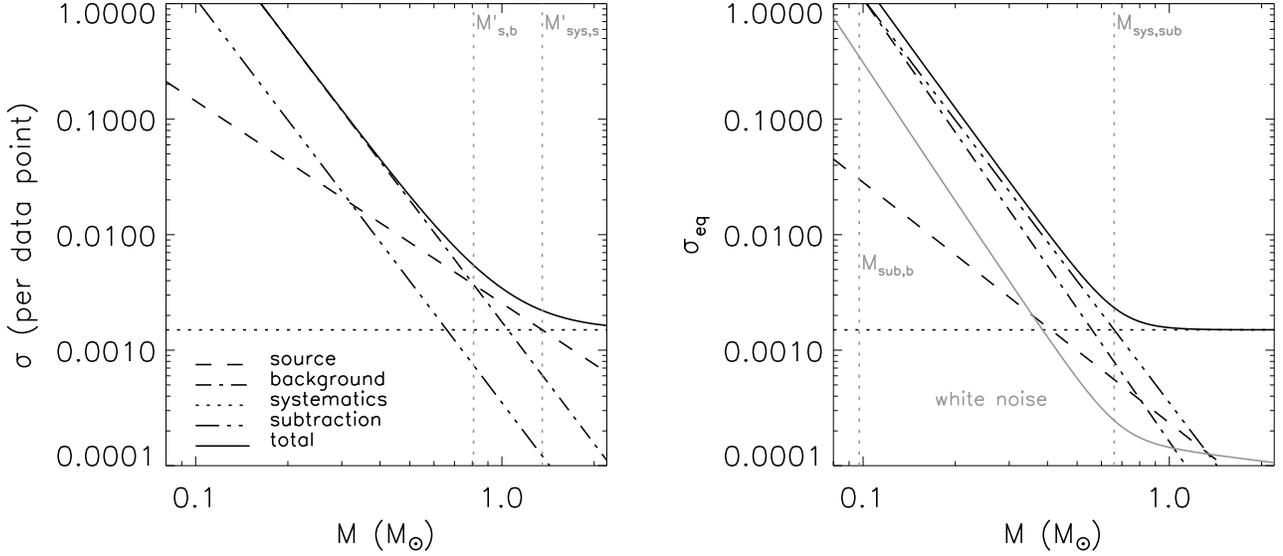,width=\linewidth}
\caption{Error budget on individual data points (left) and over a the
  duration of an equatorial transit (right) for the fixed and fiducial
  parameters of PG05a, assuming $\sigma_{\rm sys}=1.5$\,mmag and $k=0.2$.
  The black dashed, dot-dash, triple dot-dash and dotted lines show the source photon noise, background photon noise, background subtraction noise and systematics terms respectively, and the solid black line shows the total noise budget. The grey line on the right panel shows what the behaviour the total noise would have it all the components behaved as white noise.
  The grey vertical dotted lines show mark transitions between
  the different regimes, as defined in Equations~(\protect\ref{eq:msb}) to (\protect\ref{eq:msubb}).
  \label{fig:sigma}}
\end{figure*}

The form of $\sigma_{\rm eq}$, {the depth uncertainty}
for equatorial transits ($b=0$) and of the different terms that
compose it is illustrated in Figure~\ref{fig:sigma} (right
panel). Also shown for comparison is the noise level \emph{per data
  point} (left panel){, or $\mathcal{V}(1)$.}

The relative importance of the red noise components is clearly
enhanced over the transit time-scale. While the systematics term is the
same for all stellar masses, the source photon noise is a steeply
decreasing function of stellar mass, the background subtraction noise
is even steeper, and the background photon noise is the
steepest. There may thus be up to four noise regimes, starting with
the systematics-limited regime at the highest masses, followed by the
source noise-limited regime, the subtraction-limited regime, and
finally the background noise-limited regime at the lowest masses.

Equating each pair of components and solving for $M$ yields the mass
regimes in which each component dominates. This exercise was done by
PG05a to obtain $M_{\rm sky}$, the transition mass between the source
and background noise-limited regimes, which for clarity we rename
$M_{\rm s,b}$.
\begin{equation}
\label{eq:msb}
M_{\rm s,b} = C_2^{\frac{1}{\beta_X}} M_{\odot}
\end{equation}

Given the two additional noise terms that have been introduced, the
relevant transitions are now: {
\begin{equation}
\label{eq:msyss}
M_{\rm sys,s} = \left( C_4~\sigma_{\rm sys}^2
 \right)^{\frac{3}{1-3\alpha-3\beta_X}} M_\odot
\end{equation}
\begin{equation}
\label{eq:mssub}
M_{\rm s,sub}
= \left( C_4~C_5 \right)^{\frac{3}{1-3\alpha+3\beta_X}} M_\odot
\end{equation}
\begin{equation}
\label{eq:msubb}
M_{\rm sub,b}
= \left( \frac{C_4~C_5}{C_2} \right)^{\frac{3}{1-3\alpha}} M_\odot
\end{equation}}

However, it is very easy for the source noise-limited regime to
disappear altogether, because the source photon noise averages out
over the duration of the transit whereas the systematics and
background subtraction noise do not. Even if one ignores the
background subtraction term, the source-limited regime disappears if
$M_{\rm s,b} \geq M_{\rm sys,s}$.
Given the set of fixed and fiducial parameters adopted by PG05a, the
source limited regime exists only if $\sigma_{\rm sys} <
0.5$\,mmag. Adopting a more realistic value of 1.5\,mmag, there is
a direct transition between the systematics- and subtraction-limited
regime, which occurs at
\begin{equation}
\label{eq:msyssub}
M_{\rm sys,sub} = \left( \frac{C_5}{\sigma_{\rm sys}^2} \right)^{\frac{1}{2\beta_X}} M_\odot.
\end{equation} 

For the subtraction-dominated regime to exist requires $k$ to be
relatively large ($k \geq 0.2$). If this is not the case, there is a
direct transition between the systematics and background limited
regimes:{
\begin{equation}
\label{eq:msysb}
M_{\rm sys,b} =  \left( \frac{C_4~\sigma_{\rm sys}^2}{C_2}
 \right)^{\frac{3}{1-3\alpha-2\beta_X}} M_\odot.
\end{equation}}

\subsection{Detection probability ${\mathcal P}_{\rm S/N}$}

The detection probability is derived following the same method as 
PG05a, although we consider multiple, rather than single transits.

A transit observed {$N_{\rm tr}$} times is assumed to
be detectable if it gives rise to a detection statistic $\Delta
\chi_{\rm tr}^2 \geq \Delta \chi_{\rm min}^2$. If equatorial transits
of a given system are detectable, one can derive a maximum impact
parameter $b_{\rm max}$ up to which transits of such a system are also
discoverable (PG05a). This arises because $n = t_{\rm eq}
\sqrt{1-b^2} / \delta t$, where $b$ is the impact parameter of the
transit and $t_{\rm eq}$ the duration of an equatorial transit:
\begin{equation}
t_{\rm eq} = R \left( \frac{4P}{\pi GM}\right)^{\frac{1}{3}}
\end{equation}
where $R$ is the radius and $M$ the mass of the
star. {We have assumed that the planet radius $r \ll
  R$ and ignored limb-darkening}, which allows {us} to
ignore grazing transits as both extremely rare and hard to detect, and
to write $\delta = (r/R)^2$ {where $r$ is the planet
  radius.}

Therefore, the probability that transits of such a system are
detectable, i.e. that $\Delta \chi_{\rm tr}^2 \geq \Delta \chi_{\rm
  min}^2$, reduces to ${\mathcal P}_{\rm S/N} = b_{\rm max}$ when
integrated over $b$, assuming the impact parameters are uniformly
distributed between 0 and 1. However, the statement:
\begin{equation}
\Delta \chi_{\rm tr}^2 = \Delta \chi_{\rm eq} \sqrt{1-b^2},
\end{equation}
\noindent which is valid in PG05a, no longer holds here, because
$\Delta \chi_{\rm tr}^2$ is no longer simply proportional to
{$n$}. Instead, Equations~(\ref{eq:dc2s_gen}) and
(\ref{eq:sigtr}) imply: {
\begin{equation}
\label{eq:dc2tr2}
\Delta \chi_{\rm tr}^2 = N_{\rm tr}~\delta^2 \left( \frac{\sigma_{\rm w}^2}{n}
+ \sigma_{\rm r}^2 \right)^{-1} = 
N_{\rm tr}~\delta^2 \left( \frac{\sigma_{\rm w}^2}{n_{\rm eq} \sqrt{1-b^2}}
+ \sigma_{\rm r}^2 \right)^{-1}.
\end{equation}}
The expression for $b_{\rm max}$ is found by setting the left
hand side of Equation~(\ref{eq:dc2tr2}) to $\Delta\chi_{\rm min}^2$ and
solving for $b$. This yields:
{
\begin{equation}
\label{eq:psn}
{\mathcal P}_{\rm S/N} = b_{\rm max} = \sqrt{ 1 - \left[
 \frac{\sigma_{\rm w}^2}{n_{\rm eq}}
  \left(\frac{N_{\rm tr}~\delta^2}{\Delta\chi^2_{\rm min}}
      - \sigma_{\rm r}^2 \right)^{-1} \right]^2 }
\end{equation}}

Inserting the expressions for the different noise terms derived
{above gives:
\begin{equation}
\label{eq:psn2}
{\mathcal P}_{\rm S/N} = \sqrt{ 1 - \left\{   
\frac{\left( \frac{M}{M_\odot} \right)^{\frac{1}{3}-\alpha-\beta_X}
\left[ 1 + C_2 \left( \frac{M}{M_\odot} \right)^{-\beta_X}
\right]}{ C_4 \left[  \frac{N_{\rm tr}}{\Delta\chi^2_{\rm min}} 
\left( \frac{r}{R_\odot} \right)^{4}
\left( \frac{M}{M_\odot} \right)^{-4\alpha}
 - \sigma_{\rm sys}^2 - C_5 \left( \frac{M}{M_\odot} 
\right)^{-2\beta_X}\right]^{-1} } \right\}^2}.
\end{equation}}
This expression reduces, in the case of white noise only -- i.e. when
$\sigma_{\rm sys}$ {and $k$ vanish} -- to PG05a's
Equation~(15). A similar expression for $\Delta\chi^2_{\rm eq}$, the
detection statistic for equatorial transits, also ensues:
{
\begin{equation}
  \label{eq:dc2eq}
  \Delta \chi_{\rm eq}^2 = \frac{N_{\rm tr} \left( \frac{r}{R_\odot} \right)^{4} 
      \left( \frac{M}{M_\odot} \right)^{-4\alpha}}
    { C_4 \left( \frac{M}{M_\odot} \right)^{\frac{1}{3}-\alpha-\beta_X}
          \left[ 1 + C_2 \left( \frac{M}{M_\odot} \right)^{-\beta_X} \right] 
          + \sigma_{\rm sys}^2 
          + C_5 \left( \frac{M}{M_\odot} \right)^{-2\beta_X}}.
\end{equation}}
If $\Delta\chi^2_{\rm eq}<\Delta\chi^2_{\rm min}$ for a particular
star-planet system, transits of that systems are not detectable,
whatever the inclination.

\begin{figure}[t]
\centering\epsfig{file=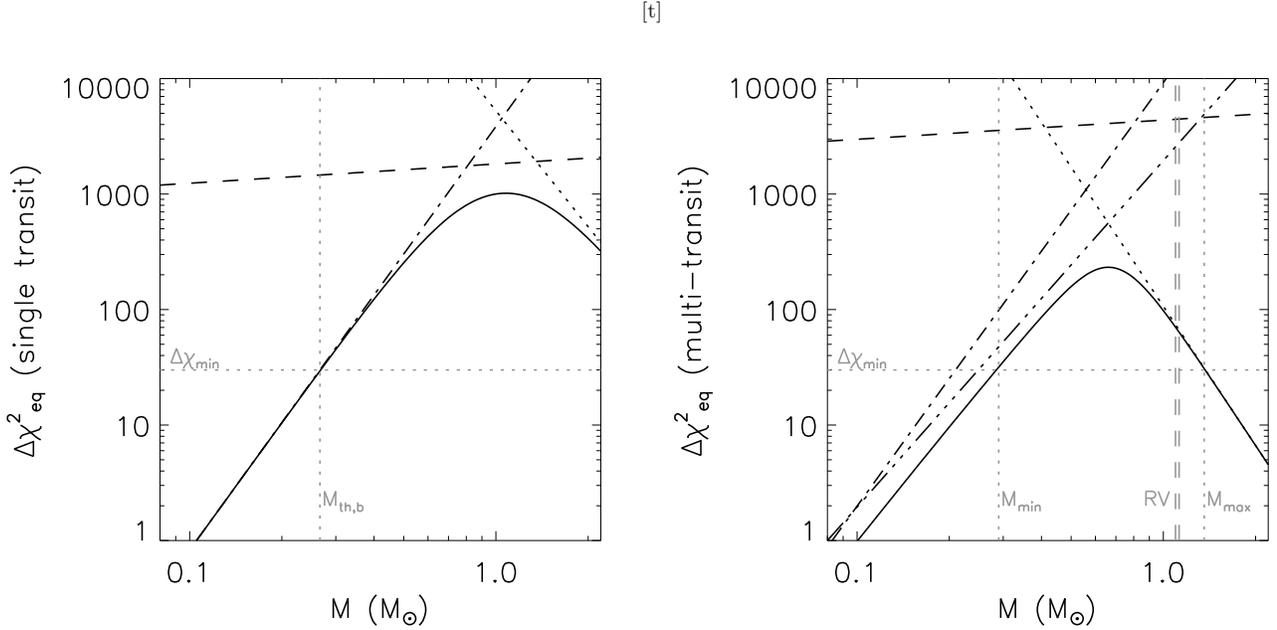,width=\linewidth}
\caption{Detection statistic $\Delta\chi^2_{\rm eq}$ for an equatorial
  transit, for individual transits and assuming the systematics are
  white (as in PG05a, left) or for multiple transits and incorporating
  both red noise terms with $\sigma_{\rm sys}=1.5$\,mmag and $k=0.2$
  (this work, right). The line styles have the same meaning as in
  Figure~\ref{fig:sigma}. The grey horizontal dotted line marks the
  detection threshold $\Delta\chi^2_{\rm min}=30$ adopted by PG05a.
  The grey vertical dotted lines mark the lower and, if applicable,
  upper mass limits between which the transits are detectable. The
  grey vertical dashed {lines mark} the mass range
  where RV follow-up is feasible with
  FLAMES$+$UVES. \label{fig:dc2eq}}
\end{figure}

The overall behaviour of $\Delta\chi^2_{\rm eq}$ as a function of mass
is illustrated in Figure~\ref{fig:dc2eq} (right panel). Also shown for
comparison is the single-transit $\Delta\chi^2_{\rm eq}$ obtained
following PG05a, i.e. assuming the systematics are white (left panel).

PG05a point out that, using $\alpha=1$ and $\beta_I=3.5$,
$\Delta\chi^2_{\rm eq} \propto M^{1/6}$ and $M^{11/3}$ in the source
and background noise-limited regimes respectively, which has the
remarkable implication that the detectability of planetary transits is
virtually independent of mass for all stars above sky,
{while} it decreases rapidly for stars below sky. In
white{, source limited} noise only, the number of
detections from a given survey is thus roughly proportional to the
number of unsaturated stars above sky. In the red noise-limited
regimes, $\Delta\chi^2_{\rm tr}$ no longer depends on
{$n$}, i.e.\ on $b$ (provided the transit is not
grazing, a given transit event contributes the same amount to the
detectability no matter what the number of observations in that
transit). Transits of a given system are thus detectable, whatever the
inclination (i.e. the transit duration), if they are deep enough
($\delta$) and enough of them are observed ({$N_{\rm
    tr}$}). Using the same values of $\alpha$ and $\beta_I$ as in
PG05a, $\Delta\chi^2_{\rm eq} \propto M^{-4}$ and $M^{3}$ in the
systematics and background-limited regimes respectively, with the
remarkable implication that transits are detectable only around stars
\emph{below} a certain mass, determined by the systematics term. The
reason is that the degree of correlation of the noise lowers the
advantage of having longer transits and lower photon noise (a larger
and brighter primary) compared to that of having deeper transits (a
smaller primary).

{The combined effect of the different noise terms
  across the entire stellar mass range is to give rise to a peak in
  $\Delta\chi_{\rm eq}^2$ versus $M$, as illustrated by
  Figure~\ref{fig:dc2eq}. This immediately points to a potentially
  very simple way of evaluating wether a given type of planet is
  detectable at all in a given cluster with a given observational
  setup: find the `peak mass', or stellar mass at which
  $\Delta\chi_{\rm eq}^2$ is maximised, by differentiating
  Equation~(\ref{eq:dc2eq}) with respect to $M$ and setting the
  derivative to zero:
  \begin{equation}
    \label{eq:solvempeak}
    C_4 \left( \frac{M}{M_{\odot}} \right)^{ \frac{1}{3} - \alpha - \beta_X } \left[
      5 \alpha + \beta_X - \frac{1}{3} + 
      C_2 \left( 5 \alpha + 2 \beta_X - \frac{1}{3} \right) 
      \left( \frac{M}{M_{\odot}} \right)^{ - \beta_X } \right] + 
    C_5 \left( 1 + 2 \beta_X \right)
    \left( \frac{M}{M_{\odot}} \right)^{ - 2 \beta_X } + \sigma^2_{\rm sys} = 0.
  \end{equation} 
  The solution of Equation~(\ref{eq:solvempeak}) could then be
  plugged back into Equation~(\ref{eq:dc2eq}), to yield
  $\Delta\chi_{\rm peak}^2$. If $\Delta\chi^2_{\rm
    peak}>\Delta\chi_{\rm min}^2$, detections are possible in the
  cluster under consideration. In that case, one can also compute the
  limits $M_{\rm low}$ and $M_{\rm up}$ of the stellar mass range over
  which detections are possible by setting the left hand side of
  Equation~(\ref{eq:dc2eq}) to $\Delta\chi_{\rm min}^2$ and solving
  for $M$. In practice, both equations cannot be solved analytically
  in the general case. If one is
  interested in calculating precise values of $\Delta\chi_{\rm
    peak}^2$, $M_{\rm low}$ and $M_{\rm p}$, the simplest way is to to
  compute $\Delta\chi_{\rm eq}^2$ for a range of $M$ and find the
  quantities of interest numerically.}

{However, as discussed in Section~\ref{sec:sigmatr},
  it is not uncommon for a single component to dominate over a
  significant portion of the mass regime. By considering dependence of
  $\Delta\chi^2_{\rm eq}$ on each of the components one at a time, one
  can obtain useful insights into what limits the transit survey's
  performance, and what mass range will be accessible.} Each of the
source, background and background subtraction terms imply a minimum
mass around which a given type of transit is detectable:
{
\begin{equation}
\label{eq:mlows}
M_{\rm low,s} = \left( C'_1 \right)^{\frac{3}{-9\alpha+3\beta_X-1}}
M_{\odot} 
\end{equation}
\begin{equation}
\label{eq:mlowb}
M_{\rm low,b} = \left( C'_1~C_2 \right)^{\frac{3}{-9\alpha+6\beta_X-1}}
M_{\odot} 
\end{equation}
\begin{equation}
\label{eq:mlowsub}
M_{\rm low,sub} = \left( C'_1~C_4~C_5 \right)^{\frac{1}{4\alpha-2\beta_X}} 
M_\odot
\end{equation}}
where {$C'_1$ is a multiple-transit equivalent of
  PG05a's $C_1$:
\begin{equation}
\label{eq:c1}
C'_1 = \frac{C_1}{N_{\rm tr}} = \frac{\Delta\chi^2_{\rm
  min}}{N_{\rm tr}~N_{{\rm s},\odot}~n_{{\rm eq},\odot}} 
\left( \frac{r}{R_\odot} \right)^{-4} 
\end{equation}}
and $M_{\rm low,s}$ and $M_{\rm low,b}$ are equivalent to PG05a's
$M_{\rm th,s}$ and $M_{\rm th,b}$. On the other hand, the systematics
term implies a maximum mass:{
\begin{equation}
\label{eq:mupsys}
M_{\rm up} = \left( \frac{1}{C'_1~C_4~\sigma_{\rm sys}^2} 
\right)^{\frac{1}{4\alpha}} M_\odot
\end{equation}}
Note that PG05a's white systematic noise term also induces an upper
mass limit detection, but it is typically larger than $3\,M_\odot$.
As long as {$M_{\rm up}$} is above the peak of the
mass function, correlated systematic noise will not significantly
affect the total number of detections in a given survey. However, it
does have the effect of preventing detections around the brightest
stars, which are arguably the most interesting, because of their
enhanced potential for follow-up.

{In general, more than one component contributes near
  the limits of the range of masses of stars around which transits of
  a given type of planet are detectable, and the expressions for
  $M_{\rm low}$ and $M_{\rm up}$ are rather complex. Again, an
  alternative is to compute each term in $\Delta\chi_{\rm eq}^2$ for
  an array of stellar masses and to find the values of $M$ between
  which $\Delta\chi_{\rm eq}^2>\Delta\chi_{\rm min}^2$.}

\section{Additional considerations}
\label{addcons}

\subsection{Turnoff mass}

Following PG05a, we compute the turnoff mass
\begin{equation}
\label{eq:mto}
M_{\rm to} = \left( \frac{\epsilon M_\odot^\beta c^2}{L_{{\rm bol},\odot}A}
\right)^{1/(\beta-1)}
\end{equation}
where $\epsilon$ is the efficiency of Hydrogen burning and $\beta$ is
the bolometric mass-luminosity index.

\subsection{Saturation mass}

Also following PG05a's expression for the number of photons at the peak
of the PSF of a given star, the saturation mass is {
\begin{equation}
\label{eq:msat}
\left( \frac{M_{\rm sat}}{M_{\odot}} \right)^{\beta_X} = \frac{4 \pi d^2}{L_{X,\odot}} 
~10^{0.4A_X}~
\left[ \left(\frac{D}{2}\right)^{-2} \frac{N_{\rm FW}}{t_{\rm exp} \pi} - S_{{\rm sky},X} \theta_{\rm pix}^2 \right]
\left\{1-\exp\left[- \ln 2 \left( \frac{\theta_{\rm pix}}{\theta_{\rm see}} \right)^2 \right]\right\}^{-1}
\end{equation}
} where $N_{\rm FW}$ is the
full-well capacity of the detector and $\theta_{\rm pix}$ is the
angular size of the pixels.

\subsection{Radial velocity follow up}

Radial velocity (RV) follow-up is necessary to confirm the planetary
nature of any detected transits and to measure companion masses. In
this section,we examine the range of stellar masses over which this is
feassible for a planet of a given mass and period.

PG05a used a fixed magnitude limit {($V=17$ or
  $V=18$)} beyond which planets were considered undetectable by the
radial velocity method. This is approximately suitable for planetary
companions to Sun-like stars: it is extremely difficult to measure
radial velocities with precisions of a few tens of m/s level beyond
{$V\sim18$} even with the largest telescopes available
at present \citep{pbm+05}. However, in cluster transit searches, many
of the detections occur around lower-mass stars, where planetary
companions may induce significantly larger radial velocity
modulations, and a more detailed treatment is needed.

For a star of a given magnitude, the minimum detectable radial
velocity amplitude $K_{\rm min}$ is highly instrument dependent, and
we examine two representative telescope / instrument combination: the
UV-Visual Echelle Spectrograph (UVES) coupled to the Fibre Large Area
Multi-Element Spectrograph (FLAMES) on the Very Large Telescope (VLT)
-- hereafter FLAMES$+$UVES -- and the High Accuracy Radial velocity
Planet Searcher on the 3.6\,m telescope at La Silla -- hereafter
HARPS.  High precision measurements tend to be limited by instrument
stability rather than by photon noise, in the sense that, if deemed
interesting enough, a given (short-period) object can be observed as
long as necessary, binning the phase-folded measurements to reduce the
photon noise contribution. However, for each telescope / instrument
combination there is also a magnitude limit {$Y_{\rm
    RV}$}, beyond which the signal-to-noise ratio achievable in a
single exposure drops below a critical level, and high precision
measurements are no-longer feasible in reasonable exposure times. As
the spectral region used typically covers the $V$ and $R$-bands,
{$Y$ should be either} $V$ or $R$, depending on which
filter the object under consideration is brightest
in. 

The radial velocity semi-amplitude induced by a given planet scales as
\begin{equation}
K \propto m P^{-1/3} M^{-2/3}
\end{equation}
where $m$ is the planet mass and we have assumed that $m \ll M$ and
that the inclination of the system is edge-on. All planets giving rise
to $K \geq K_{\rm min}$ are then assumed to be detectable around stars
with apparent magnitude down to {$Y_{\rm RV}$}, beyond
which it is assumed that high precision radial velocity measurements
are not feasible at all with the telescopes/instruments under
consideration. For HARPS, we use $Y=V$ and $Y_{\rm RV}=14$, for
FLAMES$+$UVES we use $Y=R$ and $Y_{\rm RV}=18$. Both are relatively
optimistic limits. This means that in each cluster, there is a lower
mass limit
\begin{equation}
M_{\rm RV,min} = 10^{M_{Y,\odot}-{Y}_{\rm RV}+5 \log d-5+A_Y/2.5\beta_Y} 
\end{equation}
where $Y$ is $R$ or $V$, below which no radial velocity measurements
are feasible with a given instrument, and above which the minimum
detectable planet mass is
\begin{equation}
\label{eq:mplmin}
m_{\rm min}= m_{\rm ref} \left( \frac{P}{{\rm 3~days}} \right)^{1/3}
                      \left( \frac{M}{M_{\odot}} \right)^{2/3}
\end{equation}
where $m_{\rm ref}=M_{\rm Neptune}$ for HARPS and $M_{\rm Jupiter}$
for FLAMES$+$UVES. If considering a particular planet mass $m$ across
a range of stellar masses, one can derive a maximum stellar mass
$M_{\rm RV,max}$ around which such a planet produces a detectable RV
signal by setting $M_{\rm min}$ in Equation~(\ref{eq:mplmin}) to $m$:
\begin{equation}
  M_{\rm RV,max} = 
  \left( \frac{m}{m_{\rm ref}} \right)^{3/2}
  \left( \frac{P}{{\rm 3~days}} \right)^{-1/2} ~M_{\odot} 
\end{equation}
which is independent of the {cluster properties and
  depends on the planet mass and} period only. Planets with mass $m$
and period $P$ can thus be confirmed by radial velocity with present
observational means only if they orbit stars with $M_{\rm RV,min} < M
< M_{\rm RV,max}$.

Note that, for the sake of simplicity, we have ignored a number of
important factors, including morphological differences in the spectra
of stars of different types and the impact of rotation, which broadens
the lines and degrades the radial velocity precision. 

\section{Applications}
\label{examples}

One can roughly evaluate the mass range $\left[M_{\rm min};M_{\rm
    max}\right]$ within which planets of a given radius and period in
a given cluster produce detectable transits \emph{and} RV modulations:
{
  \begin{equation}
    \label{eq:mmin}
    M_{\rm min} = {\rm max} \left( M_{\rm low}, M_{\rm RV,min} \right)
  \end{equation}
  \begin{equation}
    M_{\rm max} = {\rm min} \left( M_{\rm up}, M_{\rm to}, M_{\rm sat},M_{\rm RV,max} \right)
  \end{equation}
}

\subsection{PG05a's fiducial cluster}

Going back to the fiducial cluster of PG05a, under the relatively
optimistic assumption that $\sigma_{\rm sys}=1.5$\,mmag, the detection
of transits alone for a $1\,M_{\rm Jupiter}$ planet in a 2.5-d orbit
is possible around stars with masses between 0.28 and
$1.49\,M_{\odot}$. However, using FLAMES$+$UVES on the VLT, for which
we assume that $K_{\rm min}$ corresponds to a Jupiter mass planet in
the same orbit around the same star and that {$R_{\rm
    RV}=18$}, $M_{\rm RV,min}=1.13\,M_{\odot}$ and
$M_{\rm RV,max}=1.22\,M_{\odot}$, so that the mass range where such a
planet can be detect via transits \emph{and} radial velocity is only
$0.11\,M_{\odot}$.

The combination of correlated systematics and follow-up requirements
imposes very stringent limits on the potential of transit surveys in
open clusters. In practice, it implies an even stronger dependence
on cluster distance that illustrated in the bottom right panel of
PG05a's Figure~8.

\subsection{Example galactic open clusters}

In a subsequent paper, PG05b applied the formalism of PG05a to a
number of well-studied Galactic open clusters and, on this basis, made
the prediction that close-in Neptune- or even Earth-sized planets
should be detectable in some of these clusters via transit surveys
from ground-based 2- to 6-m class telescopes. If so, transit surveys
in open clusters might not only enable the detection of planets around
well characterised stars of known age and metallicity, but may also
lead to the first radius measurements for terrestrial planets. It is
therefore interesting to investigate the detectability of
Jupiter-sized and smaller planets in these clusters in the presence of
red noise.

\begin{table}
  \centering
\begin{tabular}{lrrrrrrlrlr}
\hline
Name & Distance & Age & Aperture & $t_{\rm exp}$ & $M_{\rm low}$ & (Cause) & $M_{\rm up}$ & (Cause) & $\Delta M$ \\
& (pc) & (Myr) & (m) & (s) & ($M_{\odot}$) & & ($M_{\odot}$) & & ($M_{\odot}$) \\
\hline
              Hyades &   46 &  625 & 1.8 &  15 & 0.15 &  (lim) & 0.55 &  (sat) & 0.47 \\
            Praesepe &  175 &  800 & 1.8 &  45 & 0.15 &  (lim) & 0.87 &  (sat) & 0.79 \\
      NGC 2682 (M67) &  783 & 4000 & 3.6 &  45 & 0.15 &  (lim) & 1.36 &   (TO) & 1.28 \\
      NGC 2168 (M35) &  912 &  180 & 3.6 &  45 & 0.15 &  (lim) & 1.49 &  (sys) & 1.41 \\
      NGC 2323 (M50) & 1000 &  130 & 3.6 &  45 & 0.15 &  (lim) & 1.49 &  (sys) & 1.41 \\
      NGC 2099 (M37) & 1513 &  580 & 3.6 &  45 & 0.15 &  (lim) & 1.49 &  (sys) & 1.41 \\
            NGC 6819 & 2500 & 2900 & 6.5 &  45 & 0.15 &  (lim) & 1.49 &  (sys) & 1.41 \\
            NGC 1245 & 2850 &  960 & 6.5 &  45 & 0.15 &  (lim) & 1.49 &  (sys) & 1.39 \\
            NGC 6791 & 4800 & 8000 & 6.5 &  45 & 0.15 &  (lim) & 1.08 &   (TO) & 0.92 \\
\hline
\end{tabular}
\caption{Masses ranges over which transits of Jupiter-sized planets in 2\,d orbits are detectable in selected Galactic open clusters, using the observational parameters of PG05b. Columns 7 and 9 give the primary cause of the upper and lower limits (sat: saturation; TO: turn-off; sys: systematics; lim: lower limit of mass range considered).\label{tab:jup_norv}}
\end{table}

We use the same test sample of 9 clusters (the Hyades, Praesepe, M67,
M35, M50, M37, NGC\,6819, NGC\,1245 and NGC\,6791) as PG05b, from
which we take the cluster parameters (distance, age, extinction) and
the observational parameters, which are similar to those of PG05a
except that the night duration is $t_{\rm
  night}=8$\,h{, the telescope apertures are 1.8\,m
  (Pan-STARRS), 3.6\,m (CFHT) and 6.5\,m (MMT) depending on the
  cluster (as selected by PG05b),} and the exposure time is $t_{\rm
  exp}=45$\,s {for all clusters except the Hyades for
  which $t_{\rm exp}=15$\,s.}

\begin{table}
  \centering
\begin{tabular}{lrrrrlrlr}
\hline
Name & Aperture & $t_{\rm exp}$ & $M_{\rm low}$ & (Cause) & $M_{\rm up}$ & (Cause) & $\Delta M$ \\
& (m) & (s) & ($M_{\odot}$) & & ($M_{\odot}$) & & ($M_{\odot}$) \\
\hline
              Hyades & 0.1 &  30 & 0.15 &  (lim) & 0.79 &  (sat) & 0.70 \\
            Praesepe & 0.1 &  30 & 0.18 &   (RV) & 1.22 &   (RV) & 1.04 \\
      NGC 2682 (M67) & 0.8 &  15 & 0.57 &   (RV) & 1.22 &   (RV) & 0.66 \\
      NGC 2168 (M35) & 1.3 &  15 & 0.64 &   (RV) & 1.22 &   (RV) & 0.59 \\
      NGC 2323 (M50) & 1.5 &  15 & 0.66 &   (RV) & 1.22 &   (RV) & 0.57 \\
      NGC 2099 (M37) & 2.7 &  15 & 0.75 &   (RV) & 1.22 &   (RV) & 0.48 \\
            NGC 6819 & 5.0 &  15 & 0.84 &   (RV) & 1.22 &   (RV) & 0.38 \\
            NGC 1245 & 6.5 &  18 & 0.93 &   (RV) & 1.22 &   (RV) & 0.30 \\
            NGC 6791 & 6.5 &  65 & 1.07 &   (RV) & 1.08 &   (TO) & 0.01 \\
\hline
\end{tabular}
\caption{Masses ranges over which transits and radial velocity modulations of Jupiter-sized planets in 2\,d orbits are detectable in selected Galactic open clusters, using SuperWASP for the Hyades and Praesepe and the observational parameters of PG05b for the other clusters, and using FLAMES$+$UVES for radial velocity follow-up. Columns 7 and 9 give the primary cause of the upper and lower limits (RV: radial velocity).\label{tab:jup}}
\end{table} 

{Table~\ref{tab:jup_norv} shows the range of stellar
  masses between which transits Jupiter-sized planets in 2\,d orbits
  produce detectable transits. For all but the most distant cluster,
  transits are detectable right down to the minimum stellar mass
  considered ($0.15\,M_{\rm odot}$, well below the limit of
  $0.3\,M_{\odot}$ adopted by PG5b), i.e.\ the addition of a
  correlated background subtraction term component does not affect the
  results. For all but the nearest clusters, the upper limit comes
  from the systematics term ($M_{\rm up}=1.49\,M_{\rm sun}$,
  independent of the cluster and observational parameters). For the
  Hyades and Praesepe, the upper limit is saturation with the
  observational setup considered here, but this can be raised by using
  shorter exposure times and / or smaller telescopes, which have the
  added advantage of providing wider fields of view. For example, the
  SuperWASP project \citet{psc+06} uses multiple 11\,cm apertures and
  has 13.5\,as pixels and an effective bandpass similar to $R$
  ($S_{\rm sky}\sim$). As it cycles between fields,
  $\delta=8$\,min. With the standard exposure times of 30\,s, we
  obtain $M_{\rm sat}=0.79\,M_{\odot}$ for the Hyades and
  $1.71\,M_{\odot}$ for Praesepe. These represent a significant gain,
  and hereafter we adopt these observational parameters for these two
  clusters. Note that one could decrease the exposure time for the
  Hyades to increase $M_{\rm sat}$ further, but this would conflict
  with the primary goal of SuperWASP, namely to search for transits
  around field stars. Overall, correlated noise does not strongly
  affect the detectability of transits of hot Jupiters in these
  clusters.}

\begin{table}
  \centering
\begin{tabular}{lrrrrlrlrlr}
\hline
Name & Aperture & $t_{\rm exp}$ & $M^{(1)}_{\rm low}$ & (Cause) & $M^{(2)}_{\rm low}$ & (Cause) & $M_{\rm up}$ & (Cause) & $\Delta M$ \\[3pt]
& (m) & (s) & ($M_{\odot}$) & & ($M_{\odot}$) & & ($M_{\odot}$) & & ($M_{\odot}$) \\
\hline
              Hyades & 0.1 &  30 & 0.08 &  (lim) &  0.28 &  (RV) & 0.50 & (sys) & 0.42 \\
            Praesepe & 1.8 &  15 & 0.08 &  (lim) &  0.63 &  (RV) & 0.52 & (sys) & 0.44 \\
      NGC 2682 (M67) & 1.8 &  15 & 0.18 & (back) &  1.06 &  (RV) & 0.52 & (sys) & 0.34 \\
      NGC 2168 (M35) & 3.6 &  15 & 0.16 & (back) &  1.19 &  (RV) & 0.52 & (sys) & 0.36 \\
      NGC 2323 (M50) & 3.6 &  15 & 0.18 & (back) &  1.25 &  (RV) & 0.52 & (sys) & 0.34 \\
      NGC 2099 (M37) & 3.6 &  15 & 0.32 & (back) &  1.49 &  (RV) & 0.49 & (sys) & 0.17 \\
            NGC 6819 & 6.5 &  15 & 0.39 & (back) &  1.70 &  (RV) & 0.46 & (sys) & 0.07 \\
\hline
\end{tabular}
\caption{Masses ranges over which transits of Neptune-sized planets in 2\,d orbits are detectable (1) and confirmable with HARPS (2) in selected Galactic open clusters, using SuperWASP for the Hyades and the observational parameters of PG05b for the other clusters. \label{tab:nep}}
\end{table}

{We now incorporate the limits induced by radial
  velocity follow-up with FLAMES$+$UVES in the calculations. The
  results are shown in Table~\ref{tab:jup}. Radial velocity or turnoff
  are now the limiting factors in almost all cases, and imply a
  stronger distance dependence of the planet yield than transits. It
  is interesting to note, however, that confirmed detections of
  transiting hot Jupiters are possible down to very low stellar masses
  in the nearest clusters.}

{We also investigate the detectability of
  Neptune-sized planets, using HARPS for radial velocity follow-up,
  still with a period of 2\,d. For such planets, the systematics term
  implies an upper mass limit of $M_{\rm up}=0.53\,M_{\odot}$
  independent of the observational set-up. We use the same
  observational setup as before except for Praesepe, where a greater
  photon-collecting capacity than SuperWASP's is needed to offset the
  smaller planet radius, so we revert to Pan-STARRS. The results are
  shown in Table~\ref{tab:nep}.  We have omitted NGC\,1245 and
  NGC\,6791 because transits of Neptune-sized planets are not
  detectable at all in these clusters. The systematics term severely
  limits the maximum stellar mass around which transits of hot
  Neptunes can be detected, while the need for radial velocity
  measurements limits the minimum mass around which they can be
  confirmed, and it is only in the Hyades that confirmed transiting
  Neptunes are expected to be detectable. We stress that these limits
  are relatively independent of theobservational setup.}

{For hot Earths, the systematics term implies a very
  stringent upper limit of $M_{\rm up}=0.13\,M_{\odot}$, and the
  formalism adopted here also precludes radial velocity confirmation
  around any stars in the clusters condidered (although it may be
  feasible to detect the radial velocity signal from a hot Earth
  around a bright star using HARPS by observing many repetitions of
  the orbit).}

\section{Conclusions}

Simple modifications have been made to the formalism of PG05a to
account for correlated noise and the need for RV follow-up. These
should lead to more realistic estimates of the efficiency of
Open Cluster transit surveys, while retaining the analytic nature of
the original formalism, which affords useful insights into the
behaviour of the detection probability as a function of mass.

Two types of correlated noise were considered: systematics, which are
constant with apparent stellar magnitude, and background subtraction
noise, which scales with the background photon noise level. The latter
behaves in a similar fashion to background photon noise itself, though
its contribution to the total noise budget on transit time-scales has a
slightly less steep dependence on the stellar mass, and therefore it
does not significantly modify the yield of a survey unless extreme
assumptions are adopted. However, the former implies a detection
probability that steeply decreases with increasing mass and therefore
curtails detections at the bright end. This effect is much stronger
than the loss of sensitivity implied by a white minimum observational
error of similar magnitude.

In the course of evaluating the impact of correlated noise on the
detectability of transits, we made a number of simplifying
assumptions, and these should be borne in mind when comparing the
predictions of the present formalism to the yield of real cluster
transit surveys. First, we have assumed that the noise budget is the
same for all stars of a given magnitude, and that every data point in
a given light curve is affected by the same noise level. In fact, both
white and correlated noise typically affect some objects and/or nights
more than others, as they depend on factors which vary from object to
object (e.g.\ crowding, position on the detector, colour) and time
(e.g.\ weather, instrumental problems). Additionally, the way we
compute the number of observed transits does not take into account the
very strong features close to integer multiples of a day that are
present in the window function of most ground-based surveys. As a
result, while the scaling laws derived here apply for the majority of
the objects in a given survey, the most significant detections in a
real survey may well occur in special cases where the time sampling
and the noise characteristics were particularly favourable.

On the other hand, the radial velocity modulation induced by the
companion in the primary, in order to measure the companion's mass, is
only detectable given present day instrumentation over a certain
{stellar} mass range which can be close to, if not
above, the maximum mass implied by the systematics term for typical
targets and observational set-ups. Thus, even though correlated
systematics may not affect the yield of Open Cluster transit surveys
significantly in terms of transit detection alone (because transits
usually remain detectable around stars close to the peak of the mass
function), it has a very serious impact on the yield in terms of
transits whose planetary nature can be confirmed and where the
companion mass can be measured. While the specific colour-magnitude
relation followed by the members of a given cluster may enable one to
exclude many of the astrophysical false positives which affect all
transit surveys without actually detecting the radial velocity
modulation of the primary, the scientific impact of any detection of a
transiting planet will be significantly lowered if its mass cannot be
measured.

To illustrate a possible application of this modified formalism, it
was applied to a selection of well-studied Galactic Open Clusters,
which were used by PG05b to show that transits of Hot Neptunes, and
even Hot Earths, should be detectable from the ground in nearby young
Open Clusters. {While correlated noise alone has
  little effect on the detectability of hot Jupiters in these
  clusters, we find that radial velocity follow-up severely limits the
  minimum mass around which their masses can be measured, which makes
  the confirmation of even Jupiter-mass planets in the more distant
  clusters difficult. Additionally, correlated systematics at the
  level of 1.5\,mmag affecting all stars in a 20 night survey imply
  that transits of hot Neptunes are only detectable around stars with
  masses below $0.5\,M_{\odot}$. For such low stellar masses, the
  planetary radial velocity signal will only be measurable in very
  nearby clusters ($<100\,pc$) with present-day facilities. If hot
  Neptunes are abundant around M-stars, some could be detected by the
  combination of small aperture, wide field surveys such as SuperWASP
  and state of the art radial velocity instruments such as HARPS.}

{The same level of systematics limits the detection of
  transits of Hot Earths to stars with masses below $0.13\,M_{\odot}$,
  irrespective of the cluster properties or observational setup. It is
  thus vital to achieve lower systematics (e.g.\ by going to space
  with CoRoT and Kepler) to detect transits of terrestrial planets,
  and particularly to detect them around stars bright enough that it
  may be possible to measure their radial velocity signal with future
  instrumentation. }

A general trend that emerges from this work is that the combination of
correlated noise and RV follow-up requirements severely limits the
choice of suitable target clusters, and effectively imposes a rather
stringent distance limit. {Additionally, we note that,
  for a given cluster, the optimal observational setup differs
  depending on the type of planet considered.}

\section*{Acknowledgements}

This work was initiated in the course of the meetings of the
International Team on Transiting Planets set up in 2005 under the
auspices of the International Space Science Institute (ISSI), and we
are indebted to all the members of this team for invigorating
discussions and useful feedback. SA gratefully acknowledges support
from a PPARC Postdoctoral Research Fellowship. This work has been done
in the context of the preparation to the CoRoT data analysis, for
which FP acknowledges support from a PRODEX
grant. {The authors are also grateful to the referee,
  Scott Gaudi, for his careful reading of the manuscript and his
  useful comments and suggestions.}

\bibliographystyle{mn2e} \bibliography{astroph.bib}

\end{document}